\documentclass[twoside,twocolumn,9pt]{article}
\usepackage{extsizes}
\usepackage[super,sort&compress,comma]{natbib} 
\usepackage[version=3]{mhchem}
\usepackage[left=1.5cm, right=1.5cm, top=1.785cm, bottom=2.0cm]{geometry}
\usepackage{balance}
\usepackage{times,mathptmx}
\usepackage{sectsty}
\usepackage{graphicx} 
\usepackage{lastpage}
\usepackage[format=plain,justification=justified,singlelinecheck=false,font={stretch=1.125,small,sf},labelfont=bf,labelsep=space]{caption}
\usepackage{float}
\usepackage{fancyhdr}
\usepackage{fnpos}
\usepackage[english]{babel}
\usepackage{array}
\usepackage{droidsans}
\usepackage{charter}
\usepackage[T1]{fontenc}
\usepackage[usenames,dvipsnames]{xcolor}
\usepackage{setspace}
\usepackage[compact]{titlesec}

\usepackage{amssymb}
\newcommand\diff{\mathrm{d}}

\usepackage{epstopdf}

\definecolor{cream}{RGB}{222,217,201}

\begin{document}

\pagestyle{fancy}
\thispagestyle{plain}
%

\makeFNbottom
\makeatletter
\renewcommand\LARGE{\@setfontsize\LARGE{15pt}{17}}
\renewcommand\Large{\@setfontsize\Large{12pt}{14}}
\renewcommand\large{\@setfontsize\large{10pt}{12}}
\renewcommand\footnotesize{\@setfontsize\footnotesize{7pt}{10}}
\makeatother

\renewcommand{\thefootnote}{\fnsymbol{footnote}}
\renewcommand\footnoterule{\vspace*{1pt}%
\color{cream}\hrule width 3.5in height 0.4pt \color{black}\vspace*{5pt}} 
\setcounter{secnumdepth}{5}

\makeatletter 
\renewcommand\@biblabel[1]{#1}            
\renewcommand\@makefntext[1]%
{\noindent\makebox[0pt][r]{\@thefnmark\,}#1}
\makeatother 
\renewcommand{\figurename}{\small{Fig.}~}
\sectionfont{\sffamily\Large}
\subsectionfont{\normalsize}
\subsubsectionfont{\bf}
\setstretch{1.125} 
\setlength{\skip\footins}{0.8cm}
\setlength{\footnotesep}{0.25cm}
\setlength{\jot}{10pt}
\titlespacing*{\section}{0pt}{4pt}{4pt}
\titlespacing*{\subsection}{0pt}{15pt}{1pt}


\makeatletter 
\newlength{\figrulesep} 
\setlength{\figrulesep}{0.5\textfloatsep} 

\newcommand{\topfigrule}{\vspace*{-1pt}%
\noindent{\color{cream}\rule[-\figrulesep]{\columnwidth}{1.5pt}} }

\newcommand{\botfigrule}{\vspace*{-2pt}%
\noindent{\color{cream}\rule[\figrulesep]{\columnwidth}{1.5pt}} }

\newcommand{\dblfigrule}{\vspace*{-1pt}%
\noindent{\color{cream}\rule[-\figrulesep]{\textwidth}{1.5pt}} }

\newcommand\remark[1]{%
  {\fboxsep=.3ex\colorbox{Purple}{\textcolor{white}{remark}}
  \textcolor{Purple}{#1}}%
}
\newcommand\wording[1]{%
  {\fboxsep=.3ex\colorbox{OliveGreen}{\textcolor{white}{wording}}
  \textcolor{OliveGreen}{#1}}%
}
\newcommand\todo[1]{%
  {\fboxsep=.3ex\colorbox{BrickRed}{\textcolor{white}{\textsc{TODO}}}
  \textcolor{BrickRed}{#1}}%
}
\newcommand\add[1]{\textcolor{blue}{#1}}
\newcommand\delete[1]{\textcolor{red}{\sout{#1}}}

\newcommand\hide[1]{\textcolor{red}{}}

\makeatother

\twocolumn[
  \begin{@twocolumnfalse}
\vspace{3cm}
\sffamily
\begin{tabular}{m{4.5cm} p{13.5cm} }

 & \noindent\LARGE{\textbf{Ideal circle microswimmers in crowded media}} \\
\vspace{0.3cm} & \vspace{0.3cm} \\

 & \noindent\large{Oleksandr Chepizhko$^{\ast}$\textit{$^{a}$} and Thomas Franosch\textit{$^{a}$ }} \\

 & \noindent\normalsize{Microswimmers are exposed in nature to crowded environments and their transport properties depend in a subtle way on the interaction with obstacles.   
Here, we investigate a model for a single ideal circle swimmer exploring a two-dimensional disordered array of impenetrable obstacles. The microswimmer moves on circular orbits in the freely accessible space and follows the surface of an obstacle for a certain time upon collision. Depending on the obstacle density and the radius of the circular orbits, the microswimmer displays either long-range transport or is localized in a finite region. We show that there are transitions from two localized states to a diffusive state each driven by an underlying static percolation transition. We determine the non-equilibrium state diagram and calculate the mean-square displacements and diffusivities by computer simulations. Close to the transition lines transport becomes subdiffusive which is rationalized as a dynamic critical phenomenon. } \\

\end{tabular}

 \end{@twocolumnfalse} \vspace{0.6cm}

  ]

\renewcommand*\rmdefault{bch}\normalfont\upshape
\rmfamily
\section*{}
\vspace{-1cm}


\footnotetext{\textit{$^{a}$}~Institut f\"ur Theoretische Physik, Universit\"at Innsbruck, Technikerstra{\ss}e 21A, A-6020 Innsbruck, Austria. }
\footnotetext{\textit{$^{*}$}~E-mail: oleksandr.chepizhko@uibk.ac.at}





\section{Introduction}




Microswimmers are  active agents able to self-propel when moving in an aqueous fluid and display non-equilibrium transport properties in striking contrast to the equilibrium dynamics of passive particles~\cite{Romanczuk2012,Elgeti2015,BechingerReview2016,Desai2017}. 
Biological microswimmers include bacteria~\cite{Lauga2006,DiLeonardo2011,Utada2014}, sperm cells~\cite{Friedrich2008,Jikeli2015,Kaupp2016,Nosrati2017}, amoebae and protozoa~\cite{Hao2015}, as well as certain algae~\cite{Kantsler2013,Jeanneret2016}, \emph{etc}. Recent technological advances have allowed fabricating artificial microswimmers, such as Janus particles~\cite{Howse2007,Buttinoni2012}, Quincke rollers~\cite{Bricard2013,Morin2017,Morin2017pre}, or 
particles with artificial flagellas~\cite{Dreyfus2005}, which have opened the possibility to study non-equilibrium phenomena in a well-defined setup. Artificial microswimmers are also expected to play a fundamental role in the nanotechnology of the 21st century, in particular, for bio-medical engineering, controlled drug delivery~\cite{KeiCheang2014,Park2017}, environmental cleansing of soil~\cite{Lien1999} and polluted water~\cite{Vilela2017,Soler2013}. 


%
Natural environments for  microswimmers are often crowded and strongly heterogeneous such that the motion  is confined to ramified structures bounded by obstacles. Then the available space typically consists of well-separated compartments weakly connected by narrow channels. A minimal theoretical framework for such crowded environments is provided by the Lorentz model~\cite{vanBeijeren1982}
where a single tracer explores a disordered environment consisting of  impenetrable randomly distributed obstacles. The tracer is confined to the void space which consists of finite pockets and, at low enough obstacle densities, of an infinite component spanning the entire system. 
At a certain critical obstacle density a percolation transition of the underlying space occurs such that the infinite component becomes fractal. This, in turn, induces anomalous transport as manifested in a subdiffusive growth of the mean-square displacement. 
So far, transport properties have been studied mostly for passive tracers, obeying Newtonian dynamics and specular scattering from the obstacles, or Brownian dynamics  in excluded volume. Although the underlying percolation transition is universal, characterized by universal critical exponents for the fractal dimension and for the growth of the static correlation length, the corresponding dynamic universality classes may depend on the microdynamics~\cite{Halperin1985,Machta1985,Hoefling2006,Horton2010,Hofling2013,Schnyder2015,Spanner2016,Mandal2017}.
Experiments and simulations on active particles in heterogeneous environments~\cite{BechingerReview2016} have revealed trapping in long-lived orbits~\cite{Chepizhko2013,Peruani2018}, rectification of transport~\cite{Kantsler2013,Guidobaldi2014,HUANG2018} and ratchet effects~\cite{Reichhardt2013b,Reichhardt2017review}, sorting of active particles~\cite{Volpe2011}, as well as  anomalous transport~\cite{Zeitz2017}. In particular, subdiffusive motion has been investigated via simulations for active Brownian particles with specular scattering~\cite{Zeitz2017} and both numerically and experimentally for Quinke rollers~\cite{Morin2017pre} in a two-dimensional Lorentz model 
in the vicinity of the percolation transition. However, the universality class appears to be same as for passive particles.

For more complex dynamic rules of the microswimmers  and their interactions with the obstacles,  new universality classes may nevertheless emerge. In particular, more realistic models for microswimmers should include the particularities of swimmers that have been unraveled in experimental and theoretical investigations. First, many biological and artificial microswimmers tend to move in circles~\cite{vanTeeffelen2008,Lauga2006,Kummel2013,Utada2014,Kurzthaler2017,Yamamoto2017} rather then in straight trajectories due to some asymmetries. 
Second, the interactions between particles and obstacles are of hydrodynamic origin and depending on the swimming mechanism may lead to an effective attraction or repulsion between swimmers and obstacles. Often the microswimmers tend to follow the boundary of obstacles~\cite{Kantsler2013,Takagi2014,Matteo2015,Spagnolie2015,Jeanneret2016,SosaHernandez2017,Ostapenko2018} rather than being merely reflected. In this regard, also the recently proposed analogy~\cite{Schirmacher2015} between magnetotransport of electrons in disordered impurity arrays in perpendicular magnetic field needs to be reconsidered.

Here,  we follow the studies performed for the magnetotransport problem~\cite{Schirmacher2015} and consider an ideal circular microswimmer in an array of randomly placed obstacles. Yet, we account for the experimental insight that the swimmers stick to the edges of the obstacles and move along the border before reorienting and thereby leaving the boundary.  
We show that this changes the dynamic properties of the transition at low densities  from an 'orbiting state' where the  swimmers orbit around isolated obstacle clusters to a 'diffusive state' where long-range transport exists. 
At high densities we find a transition to a localized state where all swimmers are confined to finite compartments.  In contrast to magnetotransport, we find an enhancement of the diffusivity upon increasing obstacle density except for the close vicinity of the localization transition. We also provide evidence that the transport properties close to the transition for these ideal circle microswimmers  are governed by  a new dynamic universality class. 

\section{Model and simulation details}
%
\begin{figure}
 \centering
 \includegraphics[scale=0.55]{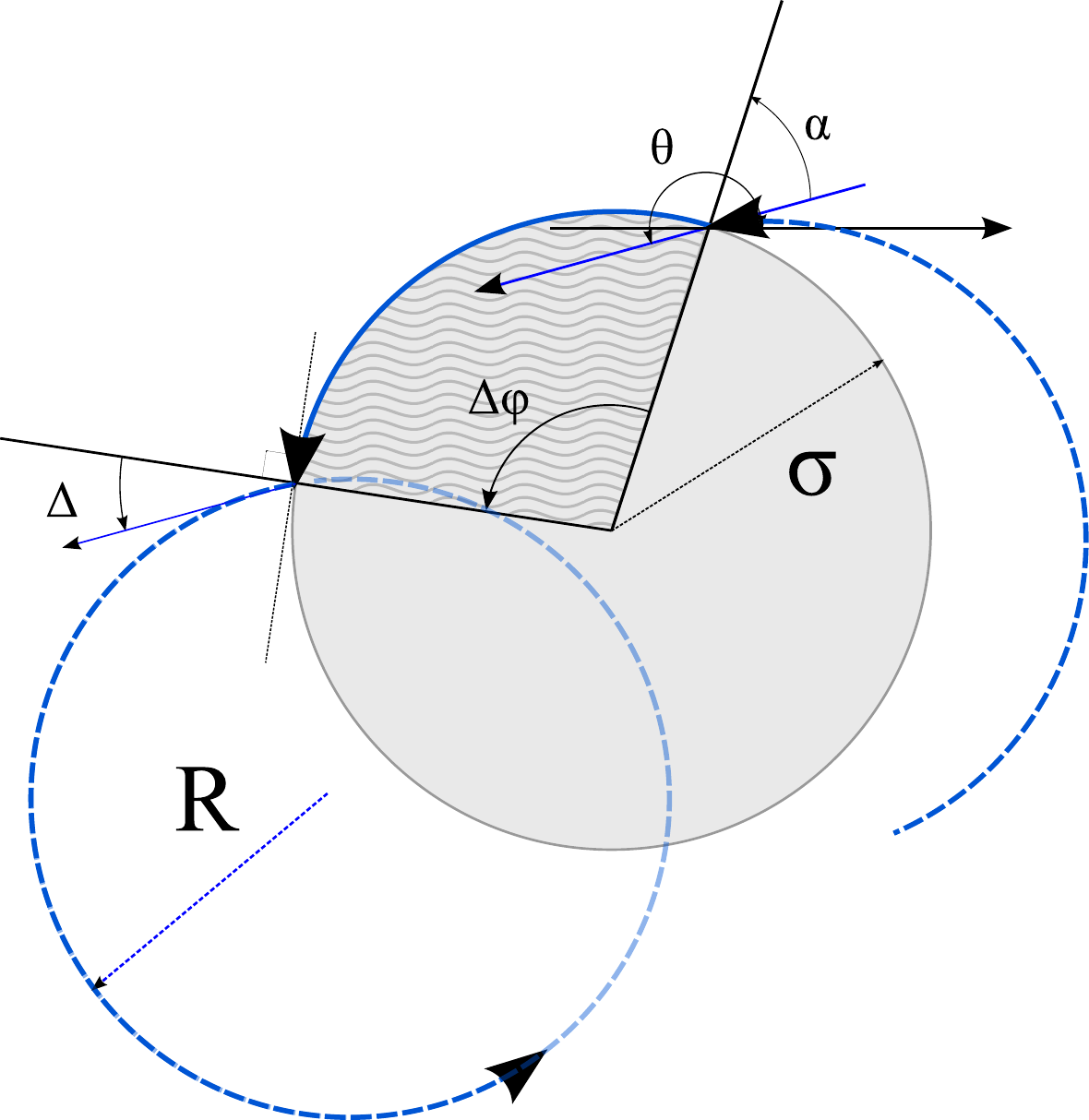}
 \caption{Illustration for particle-obstacle interaction.~The obstacle is shown as a gray circle of  radius $\sigma$. The particle trajectory is shown in blue, dashed parts represent free motion in an orbit of radius $R$, the solid part represents sliding along the obstacle surface.
 The incident angle is denoted by $\alpha$, $\Delta$ is the escape angle measured against the normal, $\Delta \phi$ represents the displacement along the surface. The direction of motion $\theta$ of the particle with respect to an axis does not change during the sliding motion, such that the two red arrows are parallel.}
\label{fig:scheme}
\end{figure}
We consider a tracer particle moving in circular orbits of radius $R$ counter-clockwise with fixed speed $v$ (corresponding to an angular velocity $v/R$)   
in a plane crowded with  randomly and independently  placed obstacles of radius $\sigma$. In particular, the obstacles may  overlap and form clusters of excluded volume. 
The obstacle radius $\sigma$ sets the unit of length and $\sigma/v$ the unit of time. The structural properties of the disorder are characterized in terms of the reduced obstacle density $n^*= N \sigma^2 /L^2$ where $N$ is the number of obstacles and $L$ the linear size of the box. 

After hitting an obstacle with an incident angle $\alpha$ the particle becomes attached to it and starts to move counter-clockwise with the same linear speed $v$ around the edge (corresponding to an angular velocity $v/\sigma$). We choose a random exit angle $\Delta \in ]-\pi/2,\pi/2[$  while keeping the initial direction $\theta$  of the velocity (with respect to a reference axis) unchanged.
 This implies that the particle follows the edge until it reaches an angle $\phi = \theta-\Delta \,(\text{mod } 2 \pi)$, see Fig.~\ref{fig:scheme}.
After escaping from the edge,  the particle again  performs a circular orbit until it hits an obstacle.    
It may happen that before escaping, the particle arrives at a `corner' where two obstacles intersect. 
In this case a new random exit angle $\Delta $ is drawn, such that  the particle follows the new obstacle as described above and the procedure is repeated. 
The interaction rule with boundaries is what essentially distinguishes our model from  magnetotransport, where tracer particles exhibit  specular reflections upon collisions with obstacles. 

Choosing a random exit angle $\Delta$  is introduced to prevent strictly or almost circular orbits, e.g. trajectories barely touching an obstacle or periodic pathways as in microswimmer billards~\cite{Spagnolie2017}.  These rare instances would give rise to long-living orbits, where the swimmer does not significantly advance and cause numerical difficulties.   
In our simulations we use $\Delta \in [-0.9 \pi/2, \,0.9 \pi/2] $ to avoid these numerical complications.
Here we choose a box size of $L=10^4\sigma$ and periodic boundary conditions, large enough to minimize finite-size effects. 
The simulations rely on an event-driven scheme and  are performed  up to  times $t=10^7$--$10^{10} \sigma/v$, depending on the combination of parameters, until the system reaches a steady state. 

Different wall following rules are conceivable, for example orientational diffusion during the wall following would change the direction of motion. Similarly, including explicitly hydrodynamic or phoretic interactions can induce rather complicated dependences of the exit angle on the incident angle. Our choice is motivated by computational simplicity while capturing the essence of microswimmer-specific dynamics.

%
In the magnetotransport problem there is an additional field-induced transition which is also of a percolative character, with a dynamic scaling scenario similar to the one for the localization transition~\cite{Kuzmany1998,Schirmacher2015}. When the density is decreased, there is a critical slowing down of transport and exactly at the transition density anomalous transport, in particular subdiffusion, occurs. 
In the diffusive regime there exists an infinite cluster (component) of the accessible void space, while at both localized states there are only isolated clusters of the void space. For densities higher than the percolation transition these clusters are formed by the overlapping obstacles. For  densities lower than the magnetic transition the particles are localized around obstacle clusters due to the magnetic field.
The transition is of a percolative character and it can be characterized by  scaling exponents, both static and  dynamic ones. The static exponents characterize the scaling properties of the underlying geometrical structure, the infinite void space cluster which is self-similar at the critical density. The dynamic exponents characterize the subdiffusion  and the suppression of the diffusivity at the critical point. The latter ones depend on the dynamic rules and the former ones purely on geometrical properties.
In the  magnetotransport problem~\cite{Schirmacher2015} it was shown that the static exponents coincide with the ones of a percolation transition, while the dynamic exponents differ drastically. As we consider here a dynamic interaction rule which is significantly different from the known ones, we expect the dynamic exponents also to be different. The determination of these exponents is a large part of the present study.

Our simulations focus on the mean-square displacements 
\begin{equation}
\delta r^2(t) := \langle [\vec{R}(t)-\vec{R}(0)]^2 \rangle, 
\end{equation}
in the stationary state, where $\vec{R}(t)$ denotes the position of the microswimmer at time $t$. The average $\langle \cdot \rangle$ includes averaging over time, different initial positions and orientations of the microswimmer, as well as over different realizations of the disorder. In particular, it includes all swimmers both trapped in finite regions as well as particles meandering through the entire system. Hence, we discuss \emph{all-cluster averages} in the following, if not  explicitly  specified otherwise.  

The simplest indicator to quantify the transport properties is the long-time diffusion coefficient 
\begin{equation}\label{eq:long_time_diffusion}
D := \lim_{t\to \infty} \frac{1}{4} \frac{\diff}{\diff t} \delta r^2 (t),
\end{equation}
and we characterize different states by finite or vanishing diffusion coefficients. 

\section{Results}
The non-equilibrium state diagram  depends on two dimensionless parameters, the reduced obstacle density $n^* = N \sigma^2 /L^2$ and the reduced orbital radius $R/\sigma$. The  state diagram displays an 'orbiting state' at low obstacle densities where the swimmers circle around isolated obstacle clusters and therefore do not display long-range transport. At intermediate densities a 'diffusive state' emerges where the swimmer can meander from one obstacle cluster to the next thereby traversing arbitrarily large distances. Last, at high densities the conventional 'localized state' arises where all swimmers are confined to finite pockets in the disordered array of obstacles.     
\begin{figure*}
 \centering
 \begin{center}
 \includegraphics[angle=90,scale=0.45]{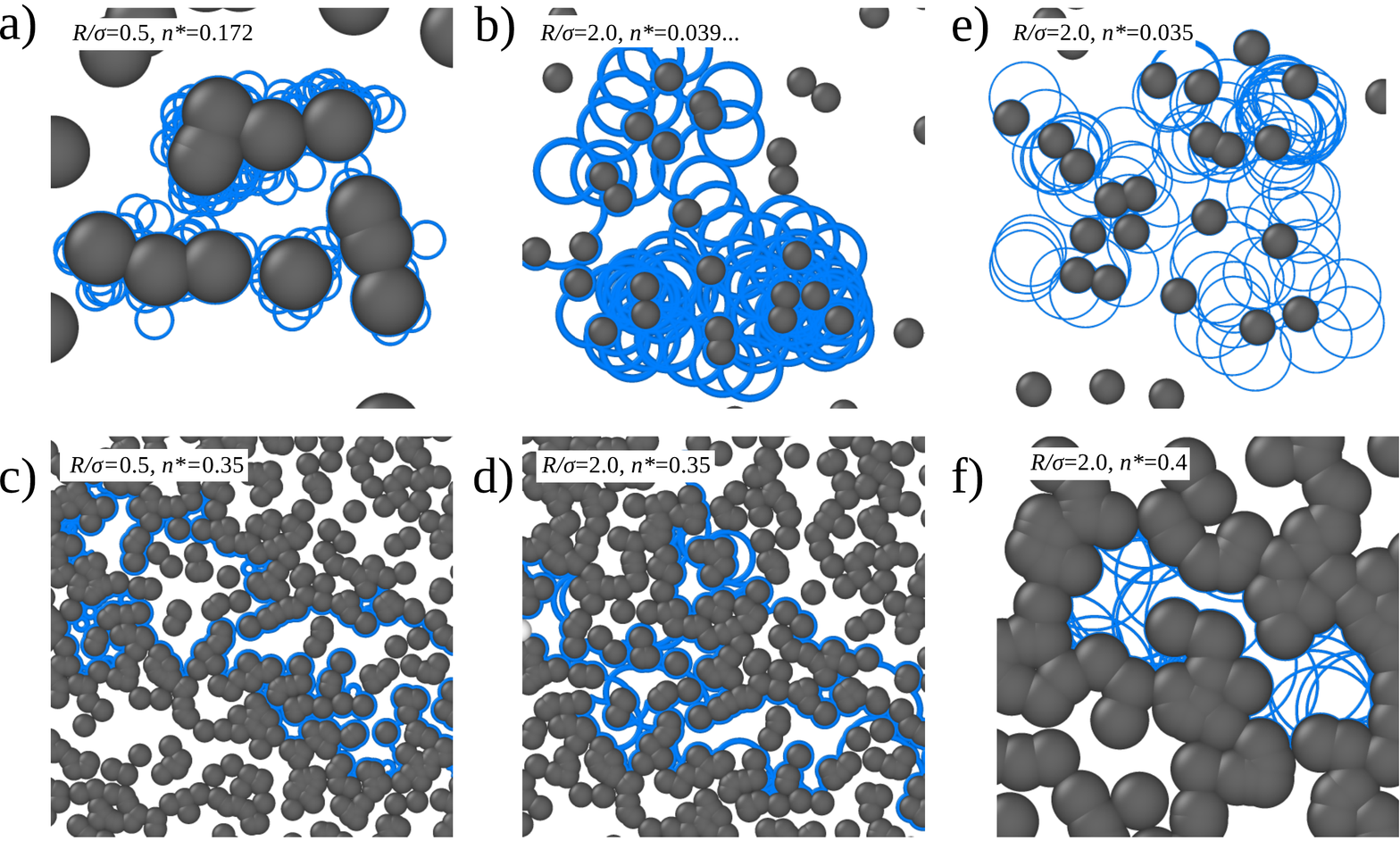} 
\end{center}
 \caption{Illustrative snapshots of trajectories of a circle swimmer in a disordered environment.
Panels (a), (c) are for orbital radius $R=0.5\sigma$, (b)  (d)  for $R=2.0\sigma$. (a) Trajectory for  density $n^*=0.172$, close to the meandering transition $n^*_m(R=0.5)\approx 0.159592$. (b) Trajectory almost at the corresponding transition density $n^*_m(R=2.0)\approx 0.039898$. (c), (d)  trajectories  at high density, $n^*=0.35$, close to the percolation transition value $n_c$. The configuration of obstacles and starting point is the same for both cases. (e)  orbiting state at $R/\sigma=2.0$ and $n^*=0.035<n_m^*(R/\sigma)$. (f)  localized state at $R/\sigma=2.0$ and $n^*=0.4>n^*_c$.}
 \label{fig:ill_snapshot}
\end{figure*}

Typical trajectories in the three respective states are displayed in Fig.~\ref{fig:ill_snapshot}. One infers that it is more difficult for a particle to 'jump' from one obstacle cluster to another for small orbital radii [Fig.\ref{fig:ill_snapshot}(a)] as the positions for a successful jump are limited and might be skipped when the particle follows the wall. At high densities, in the vicinity of the 'localization transition' [Fig.\ref{fig:ill_snapshot}(c),(d)], the interaction with the obstacles amplifies transport providing means for a quick escape from pockets formed by obstacles. Below a 'meandering transition', particles orbit around finite obstacle cluster without long-range transport [Fig.\ref{fig:ill_snapshot}(e)]. 
Last, above the localization transition all orbits are confined to finite pockets [Fig.\ref{fig:ill_snapshot}(f)].

\subsection{Non-equilibrium state diagram and diffusivity map}

\begin{figure}[htp!]
 \centering
 \begin{center}
  \includegraphics[scale=0.35]{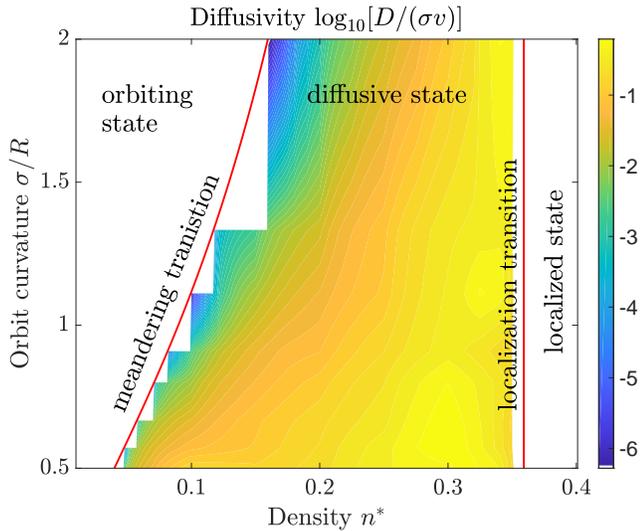} 
 \end{center}
  \caption{Diffusivity map  in terms of the  reduced density $n^*$ and the inverse reduced orbital radius $\sigma/R$. The color coding is in decadic logarithms ($\log_{10}$) of the diffusivity $D$ in units of $\sigma v$. The line for the meandering transition is defined by Eq.~\eqref{eq:nm_vs_r}. Orbiting state means that particle `orbits' around isolated obstacle clusters and the trajectory does not span the whole system. Diffusive state means that particle can explore all available space. In the localized state the microswimmer is trapped in pockets formed by overlapping obstacles.}
  \label{fig:colorful_state_diagram}
\end{figure}

We have extracted diffusivities for various obstacle densities and orbital radii from the simulations and constructed a color map for the diffusivities [Fig~\ref{fig:colorful_state_diagram}], which includes the transition lines to states of vanishing diffusivity. The highest values of the diffusion coefficients are found for fairly high obstacle densities $n^* \simeq 0.3$ quite close to the localization transition. This is rather distinct from the corresponding state diagram of magnetotransport, where the maximum of the diffusivity is found at intermediate densities~\cite{Schirmacher2015,Siboni2018}. Therefore we  infer already here that the detailed rules of interactions with the boundaries do have a significant impact on the transport properties. 

Nevertheless, the localization transition in the state diagram at high densities coincides with the geometric percolation transition of the void space as for all Lorentz models for single particle transport. In particular, the transition line is independent of the orbital radius. The critical reduced density is well known  $n^*_c = 0.359081\ldots$~\cite{Hoefling2006,Hoefling2007prl,Bauer2010, Schirmacher2015}. 
For this transition line, it is irrelevant whether the swimmers move in circular orbits and how they interact with the hard boundaries, since all swimmers are confined to finite pockets whose size is determined by the distribution of obstacles only. 

The second 'meandering transition' from the diffusive state to the orbiting state is anticipated to depend on the detailed rules of how the swimmer escapes from one obstacle cluster to the next. For magnetotransport, where the particles undergo specular scattering, it has been shown that the  location of the transition line is provided by the  percolation of obstacles with an effective radius~\cite{Kuzmany1998,Schirmacher2015} $\sigma + R$. Thus the space traversed by the particles consists of the 'halos' around the obstacles, and it is sufficient to decide if  there is  a connecting path through
the entire system via these halos. Implicitly, this picture relies on ergodicity, i.e the particle explores all accessible regions of the halo structure and finds eventually the connecting path to the next obstacle cluster. 
Then the  critical density for the meandering transition can  be related to the percolation density~\cite{Kuzmany1998}
\begin{equation}
  n^*_m(\sigma,R) = n^*_c \frac{\sigma^2}{(\sigma+R)^2} = 0.359081\ldots \frac{\sigma^2}{(\sigma+R)^2} \,.
  \label{eq:nm_vs_r}
\end{equation}
This relation is justified by the fact that transport can only happen if the microswimmers are able to jump from one obstacle cluster to another. A necessary condition is that  the disks comprised of obstacles and the halos of radius $\sigma+R$ percolate~\cite{Kuzmany1998,Schirmacher2015}.
For our circle microswimmers we find that the transition line is identical to the one predicted for magnetotransport, hence the condition is also sufficient. Thus, although the dynamic rules to interact with the obstacles are changed drastically, the ergodicity still holds and the microswimmer enters the connecting paths. Let us emphasize, that for the state diagram it is  irrelevant, how long it takes for the swimmer to find the connecting region, yet, this will be important for the diffusivities and their behavior close to the transition lines.

\begin{figure} 
 \centering
 \begin{center}
  \includegraphics[scale=0.9]{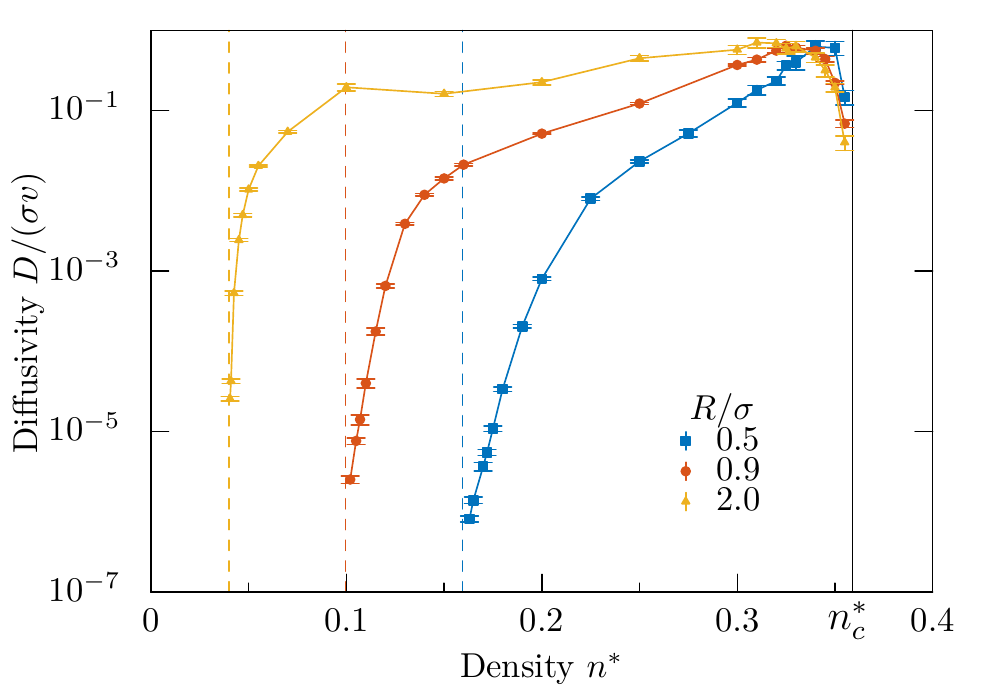} 
 \end{center}
  \caption{Diffusion coefficient $D$ as function of obstacle density  $n^*$ for different  orbital  radii $R$. 
In general, the diffusivity grows with the density of obstacles $n^*$ up to the vicinity of the percolation transition. For rather small interval of values of density just before the transition, the diffusion coefficient quickly drops down. The critical density, $n^*_c=0.359081$ is shown as a black line.}
  \label{fig:D_vs_n_star_new}
\end{figure}

The evolution of the diffusion coefficient with obstacle density is exemplified in Fig.~\ref{fig:D_vs_n_star_new} for selected orbital radii. 
The maximal diffusivities are found close to the localization transition, in striking contrast to magnetotransport, where it has been argued that the maximal diffusivity occurs at the geometric mean of the two transition densities at the same orbital radius~\cite{Siboni2018}. For microswimmers it appears that these maxima are shifted to higher densities, which we refer to as \emph{crowding-enhanced transport}.  By following the boundaries before detaching and undergoing a circular motion again, the swimmer covers larger distances at higher obstacle densities. Reversely, to benefit from this mechanism the density of the obstacles has to be higher than in magnetotransport. Obviously crowding-enhanced transport works only below the localization transition, therefore the densities have to be not too large. 

A new feature emerges at large orbital radii $R/\sigma \gtrsim 2$, where two maxima emerge upon increasing the obstacle density [Fig.~\ref{fig:D_vs_n_star_new}]. We rationalize this finding in terms of a competition of the conventional orbiting transport and the crowding enhanced transport due to following the boundaries. Correspondingly there is a minimum in between where the swimmer cannot benefit either from the orbiting nor the crowding enhanced transport.  

\subsection{Mean-square displacements and critical behavior}
More quantitative information is provided by the mean-square displacement (MSD)  $\delta r^2(t) = \langle [\vec{R}(t)-\vec{R}(0)]^2 \rangle$. In particular, the MSD varies drastically as the transition lines are approached and critical slowing-down of the dynamics is expected. Here we focus on the meandering transition and elucidate the critical properties.

\begin{figure*}[t!]
\centering
\includegraphics[scale=0.8]{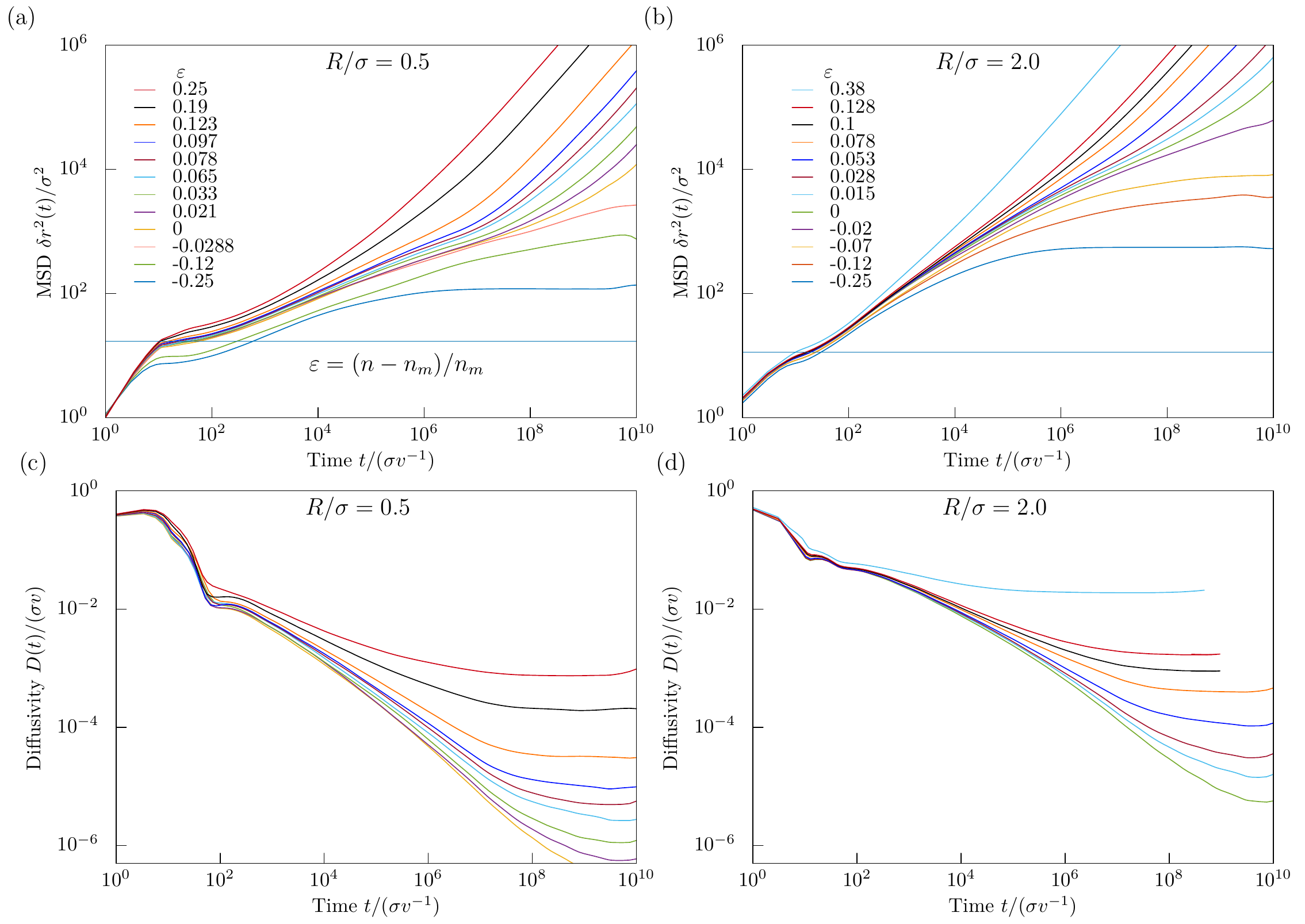}
\caption{Panels (a) and (b): Mean-square displacement $\delta r^2(t)$ vs. time for (a) $R/\sigma=0.5$ with the transition density $n_m^*(0.5) = 0.15959$ and (b)  $R/\sigma=2.0$ with $n_m^*(2.0)=0.039898$. Density decreases from top to bottom in terms of the separation parameter $\epsilon = (n-n_m)/n_m$.  
The panels (c) and (d) represent the time dependence of the diffusivity $D(t)$ in the meandering state $\epsilon \geq 0$ for $R/\sigma=0.5$ and $R/\sigma=2.0$ respectively. The blue lines in panels (a) and (b) represent the values for estimated local maxima of the MSD (when initial ballistic regime changes to subdiffusive) at respective values of critical densities. Estimation is desrcibed in the main text.}
\label{fig:msd_and_D_vs_time_R2p0}
\end{figure*}

For small times the MSD displays persistent swimming motion $\delta r^2(t) = v^2 t^2$ [Fig.~\ref{fig:msd_and_D_vs_time_R2p0}(a,b)] which holds up to times $t \lesssim \text{min}(\sigma/v, R/v)$ where the curvature of the orbits or the collisions with the walls can be ignored. 
For certain densities the MSD displays a maximum at intermediate times $t \simeq 10 \sigma v^{-1}$  and intermediate length scales $\delta r^2 \simeq (4.5 \sigma)^2$, ($\delta r^2 = c ( R + \langle N_c \rangle \sigma)^2$ where $\langle N_c \rangle$ is the average number of particles in a cluster.  This non-monotonic behavior reflects the orbiting around small clusters of obstacles [compare Fig.~\ref{fig:ill_snapshot}(a),(b)]. We estimate the local maximum to be proportional to the cluster perimeter squared, the perimeter being  proportional to the average number of obstacles in a cluster. We measure  the phenomenological coefficient of proportionality to be around $c\approx 0.9$. The estimated values for these local maxima are  shown as blue horizontal lines in Fig.~\ref{fig:msd_and_D_vs_time_R2p0}(a,b) respectively.  Our estimated values intercept roughly the MSD curves for $n^*_m(R)$ at the local maxima. 

Then a window of subdiffusive transport becomes apparent until an obstacle-density-dependent crossover to either a linear increase or a saturated MSD is reached. The linearly increasing MSD occurs in the diffusive state only, however, as all-cluster averages are considered, only a small fraction of tracer particles display long-range transport. Therefore we refer here to heterogeneous diffusion, a superposition of localized particles and particles meandering through the entire system. In contrast, for densities below the meandering transition, $n^* < n^*_m(R)$, all particles orbit around finite clusters of obstacles  and no long-range transport occurs. The window of subdiffusion grows as the transition density is approached indicative of critical behavior.

An equivalent quantity to the MSD is the time-dependent diffusion coefficient
\begin{equation}
 D(t) := \frac{1}{4}  \frac{\diff}{\diff t} \delta r^2(t)\,,
\end{equation}
which highlights the suppression of transport as the meandering transition is approached. As can be inferred from  
Fig.~\ref{fig:msd_and_D_vs_time_R2p0}, the time-dependent diffusion coefficient slows down for times larger than a few orbiting periods by orders of magnitude upon diluting the obstacle density. The curves display data collapse for intermediate times but fan out to saturated values for very long times. The crossover time scale from subdiffusive transport to heterogeneous diffusion exceeds microscopic times by several decades and signals critical slowing-down of the dynamics. In the orbiting state [not shown] the time-dependent diffusion coefficient rapidly goes to zero for times larger than the corresponding crossover time. 

\begin{figure*}
 \centering
 \includegraphics[scale=1.00]{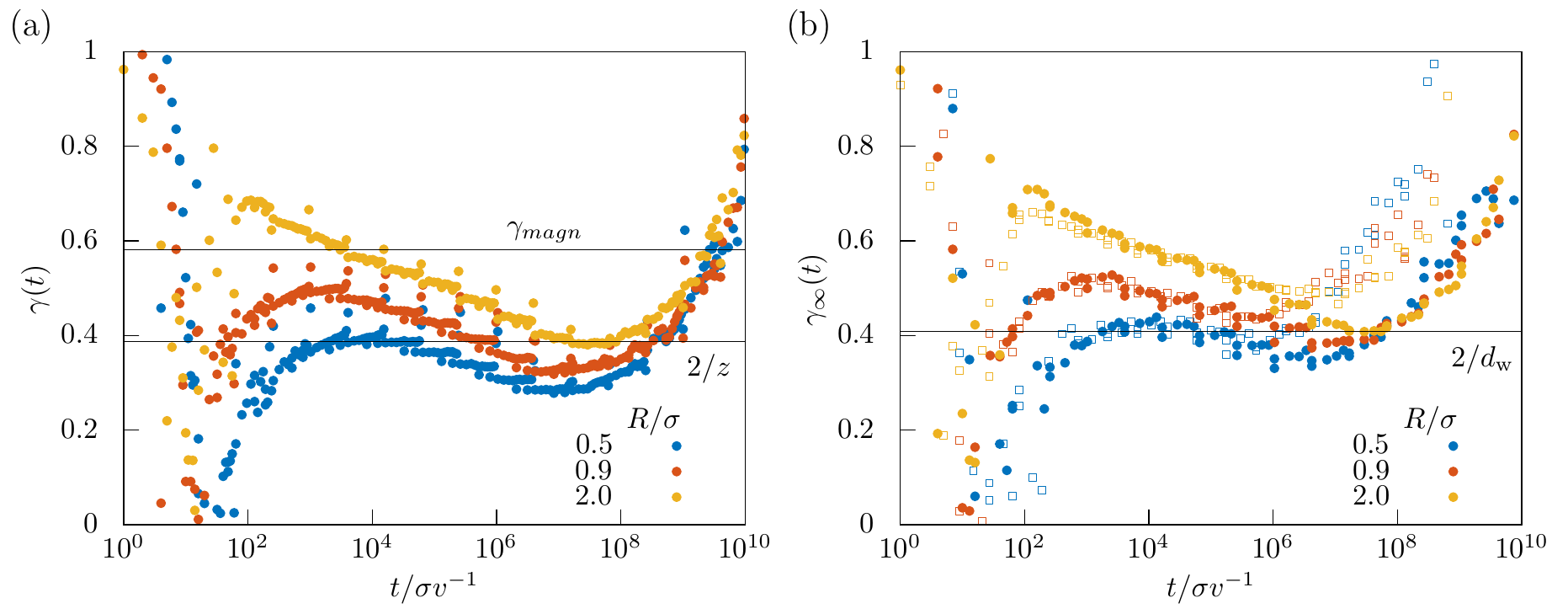} 
 \caption{(a) Time-dependent local exponent $\gamma(t)$  for three different values of $R/\sigma$. The critical value for magnetotransport~\cite{Schirmacher2015} $\gamma_{\text{magn}}=0.581$   is given as a black horizontal line as guide to eye.  
 (b) Time-dependent local exponent for the transport on the infinite cluster $\gamma_\infty(t)$
 for different values of reduced radii $R/\sigma$. Filled symbols show data for simulation run times $T = 10^{10} \sigma/v$, open symbols for shorter runs $T=10^9 \sigma/v$. }
 \label{fig:meandering_transition}
\end{figure*}

\begin{figure}
 \centering
 \includegraphics[scale=0.75]{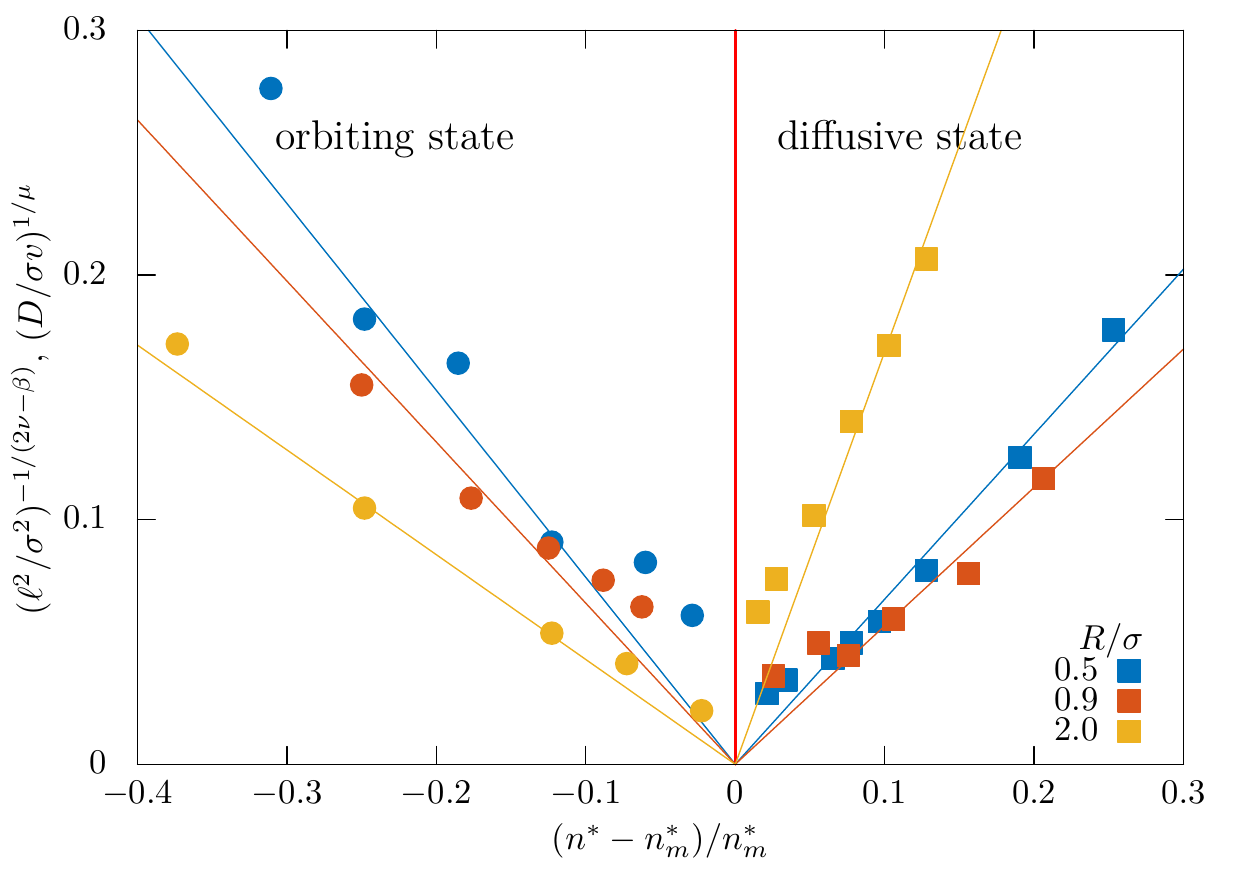} 
 \caption{Rectification of the localization lengths $\ell$ (circles, negative side),  and diffusivities $D$ (squares, positive side) for different radii.  
  }
 \label{fig:l_and_d}
\end{figure}

The anomalous diffusion at the meandering transition can be monitored conveniently in terms of the local exponent of the mean-square displacement
\begin{equation}
 \gamma(t) :=  \frac{\diff\, \log (\delta r^2(t))}{\diff\, \log(t)}\, ,
\end{equation}
as shown in Fig.~\ref{fig:meandering_transition}(a). The local exponent at the meandering transition is connected to the dynamic exponent $z:= \lim_{t \rightarrow \infty} 2/\gamma(t)$ as it is expected that 
\begin{equation}
\delta r^2 (t) \propto t^{2/z} , \qquad \text{for } t\to\infty .
\end{equation}
The local exponent close to the meandering transition remains significantly below unity for several decades in time and even below the suggested exponent for magnetotransport~\cite{Schirmacher2015} $\gamma_{\text{magn}}=2/z_{\text{magn}} \approx 0.581\dots$, where $z_{\text{magn}} \approx 3.44\dots$ . Thus the interaction rule with the boundaries now strongly slows down transport and is a relevant perturbation for the dynamic universality class, which is one of the most important results of this work. Yet, for the time windows considered, there is no clear plateau emerging and it is hard to extract a dynamic exponent characterizing the anomalous increase of the mean-square displacement. Furthermore, the minima arising in the local exponent do not coincide, prohibiting to draw a solid conclusion on a universal exponent for circle swimmers in crowded environments.   

Better statistics for  anomalous transport can be collected by restricting the swimmers to the infinite cluster only. The corresponding mean-square displacement is
\begin{equation}
\delta r_\infty^2(t) := \langle [ \vec{R}(t)-\vec{R}(0)]^2 \rangle_\infty \,,
\end{equation}
where the brackets $\langle \cdot \rangle_\infty$ indicate an average over  swimmers on the infinite cluster only.  
We  collect all trajectories at the critical density $n_m(R)$ where the individual mean-square displacement does not saturate in the simulation time and take their mean to estimate $\delta r^2_\infty(t)$. 
The expectation is that transport on the infinite cluster at the critical point becomes also anomalous
\begin{equation}
	\delta r^2_{\infty} (t) \propto t^{2/d_\text{w}} , \qquad \text{for } t\to\infty, 
    \label{eq:approaching_dw}
\end{equation}
with a critical exponent $d_{\text{w}}$ referred to as walk dimension~\cite{Gefen1983,Havlin1987,benAvraham2000}.
By scaling arguments one can connect the walk dimension with the dynamic exponent by the relation 
\begin{equation}
z  =  \frac{2 d_{\text{w}}}{2+ d_{\text{f}} - d} ,
\label{eq:z_scaling_relation}
\end{equation}
where $d_{\text{f}}$ is the fractal dimension~\cite{Grassberger1999} of the infinite cluster at the critical point and $d$ is the spatial dimension. 
We have extracted the local exponent for the infinite-cluster dynamics 
\begin{equation}
\gamma_\infty(t)  := \frac{\diff \log (\delta r^2_\infty(t))}{\diff \log(t)},
\end{equation}
 for three  orbital radii [Fig.~\ref{fig:meandering_transition}(b)]. The data show a trend to convergence to a single value up to a certain time after which they start to increase again. This increase is attributed to finite simulation times, since for shorter simulations this increase is shifted to earlier times [Fig.~\ref{fig:meandering_transition}(b)]. Hence we conclude that the scenario of a single critical exponent $2/d_{\text{w}} := \lim_{t\to\infty} \gamma_\infty(t)$ is consistent with the data. Our best estimate is $d_\text{w}=4.90 \pm 0.45$ which yields together with the known fractal dimension $d_{\text{f}} = 91/48$ in 2D and Eq.~\eqref{eq:z_scaling_relation} the estimate for the dynamic exponent $z = 5.17 \pm 0.48$, also consistent with the all-cluster-averaged data [Fig.~\ref{fig:meandering_transition}(a)]. 

Next we corroborate that the meandering transition is governed by an underlying percolation transition. The critical properties of the geometric transition are governed by two independent exponents.  One may be taken as the fractal dimension $d_{\text{f}}$  of the infinite cluster at criticality. The second exponent $\beta$ characterizes how the  weight of the infinite cluster is expected to  grow  $ \sim \varepsilon^\beta$ where $\varepsilon=(n^*-n^*_m)/n^*_m$ is the reduced distance to this transition point.
The linear size of the largest finite cluster grows as $\xi \sim |\varepsilon|^{-\nu}$ and by standard scaling arguments~\cite{Stauffer_book} 
one derives the exponent relation $\beta/\nu = d- d_{\text{f}}$. The average square cluster size $\ell^2$  then is anticipated to diverge as 
\begin{equation}
\ell^2 \sim (-\varepsilon)^{-(2\nu-\beta)} , \qquad \text{for }  \varepsilon \to  0 , \quad \varepsilon < 0,
\end{equation} 
upon approaching the transition from the orbiting state. In percolative transport $\ell$ plays the role of a localization length and it be extracted directly from the saturation of the MSD 
\begin{equation}
 \ell^2 = \lim_{t\rightarrow\infty} \delta r^2(t), 
\end{equation}
assuming that the swimmers sample the finite clusters uniformly. The critical exponents for 2D percolation are known exactly~\cite{Havlin1987} 
$\nu = 4/3, \beta = 5/36$. 

Inspection of  Fig.~\ref{fig:msd_and_D_vs_time_R2p0} shows that the localization length diverges upon approaching the transition. We have extracted the localization lengths by extrapolating the MSD to infinite times and rectified the data such that the theoretical prediction yields a straight line  $\ell^{2/(2\nu-\beta)} \sim \varepsilon$ for $\varepsilon \to 0, \varepsilon < 0$.  The result is shown and consistent for two radii, $R=0.5$ and $R=2.0$.

In the diffusive state  the time-dependent diffusion coefficients extrapolate to finite values at long times [Fig.~\ref{fig:msd_and_D_vs_time_R2p0}]. Approaching the meandering transition the diffusion coefficients decrease by orders of magnitude. From dynamical scaling~\cite{Kertesz1984} a power-law critical behavior
\begin{equation}
  D \sim \varepsilon^\mu\,, \qquad \varepsilon \to 0, \quad \varepsilon > 0 ,
\end{equation}
is anticipated. The conductivity exponent $\mu$ should be related to the dynamic exponent $z$ via~\cite{benAvraham2000} 
\begin{equation}
\mu = (z-2) (\nu-\beta/2) \approx 4.0 \pm 0.38 \,.
\end{equation}
This value is significantly higher than the known conductivity exponent of random resistor networks for percolative transport on 2D lattices~\cite{Grassberger1999,Kammerer2008}, $\mu_{\text{lat}}=1.310$, as well as the one  for  magnetotransport~\cite{Schirmacher2015}, $\mu_{\text{magn}}=1.82$. We  rectify the data for the diffusion coefficients by plotting  $D^{1/\mu}$ vs. $\varepsilon$, the theoretical expectation is then a straight line. Our data are consistent with this prediction for the different radii considered [Fig.~\ref{fig:l_and_d}], yet there remain large numerical uncertainties.

\section{Discussion and Conclusion}

We have investigated the dynamics of an ideal circular microswimmer in a crowded environment and elaborated the non-equilibrium state diagram as well as the critical behavior close to the meandering transition.  
Our model incorporates the experimental insight that microswimmers tend to follow the boundaries for a certain time before entering the bulk again. This new mechanism has drastic ramifications for the transport properties although the non-equilibrium state diagram remains unaffected. The diffusivities become maximal close to the localization transition which entails crowding-enhanced transport. 

The model also displays a meandering transition from an orbiting to a diffusive state upon increasing the obstacle density. This transition corresponds to a critical phenomenon with an underlying geometric percolation transition, yet, the dynamics is strongly slowed down due to the following of the boundaries. We have extracted the dynamic critical exponents and corroborated the dynamic scaling scenario by relating the anomalous increase of the mean-square displacement to the vanishing of the diffusion coefficient as the transition is approached. The values of the dynamic critical exponents differ significantly from the known ones for random resistor networks or magnetotransport and we conclude that our model belongs to a new universality class. Universality has been tested by considering different values of the orbiting radius.

It is interesting to ask how a new universality class can emerge in percolative transport. The standard picture is that one can map the structure to an electric network that encodes the same critical dynamics as the original model~\cite{benAvraham2000}. Then initially the links display a distribution of conductances $\rho(\Gamma)$, while  after coarse-graining all connections behave identically. Thus, this distribution renormalizes to  a delta function, and the system belongs to the same universality class as the random resistor network. Yet, if the initial distribution displays a power-law tail $\rho(\Gamma) \sim \Gamma^{-\alpha}, 0 < \alpha < 1$ \ for weak conductances, $\Gamma \to 0$, the weak links may dominate the transport properties of the system~\cite{Straley1982}.  In this case the power-law tail persists even upon coarse-graining and the entire distribution becomes self-similar. A hyperscaling relation separates the two cases and the conductivity exponent is determined by~\cite{Straley1982,Stenull2001} 
\begin{equation}
\mu = \text{max}[ \mu_\text{lat}, (d-2)\nu +1/(1-\alpha)  ],
\end{equation}
where $d$ is the spatial dimension of the surrounding space and $\mu_{\text{lat}}$ is the universal conductivity exponent of the random resistor network. Reversing the argument for our case $(d=2)$, yields that there are weak links in the transport problem characterized by an exponent 
\begin{equation}
	\alpha = 1-\frac{1}{\mu} = 0.75 \pm 0.07 \approx \frac{3}{4}\,.
	\label{eq:aexponent}
\end{equation}
The power-law tail in the conductivities reflects that a particle may follow the boundary  of an obstacle cluster for a long time before finally jumping to a neighboring one. Thus, there are narrow passages connecting the clusters and the swimmer has to find the right transition pathways. The mechanism of following the boundaries rather than specular scattering in magnetotransport appears to render the power-law tails even steeper.  

The simple fraction value for $\alpha \approx 3/4$ is deceptive, since its value is sensitive to the large value of the conductivity exponent $\mu \approx 4.0$. In particular, the large value implies that the critical region is small, the diffusion coefficient decreases drastically within a rather narrow interval close to the meandering transition. This is compatible with our observation that achieving convergence and reaching the critical region is extremely demanding. From our simulation data there remains a significant uncertainty for the values of the dynamic critical exponent, which affects the value $\alpha$. 

A natural extension of the model is to include the orientational or translational  diffusion of the circle swimmer while orbiting in bulk. 
Clearly, the sharp meandering transition will be lost, since the swimmer can explore the entire system without relying on jumping from one obstacle cluster to another. Yet, if the diffusion coefficients are small, transport is anticipated to be strongly suppressed for low densities, while the behavior in the diffusive state will be changed only slightly. To make comparison to experimental realizations  these diffusive processes have to be accounted for properly.

Our model constitutes the first step  to describe relevant biological systems.  
Different regimes of the model regarding the ratio of orbit radius and obstacle size, as well as the density of obstacles can be connected to various biological processes, that occur on micro- to mesoscopic scales. 
Then the  high-density regime may  be linked to the transport of the microorganisms through  porous media or rough channels. In particular, the results can be explored in connection to the problem of sperm navigation, guidance, in the channels. Recently, the problem was approached theoretically and experimentally~\cite{Elgeti2010biophys,Jikeli2015,Nosrati2017,HUANG2018}, and is important for different medical aspects including precise drug delivery~\cite{Xu2018}.
Similarly, the low-density regime might be applied to animals migrating on vast planes with natural obstacles, such as trees, rocks, or predators.

Further extensions of the model may  take into account the interactions among microswimmers. The simplest way is to consider them as hard disks or interacting via short-range repulsion. This might lead to clogging~\cite{Zuriguel2014,Reichhardt2018,Peter2018} and other additional obstructions of transport. Similarly, incorporating  an external driving force \cite{Leitmann2013,Benichou2014,Benichou2018,Reichhardt2018JPCM} might result in additional enhancement or decrease of diffusivity  as negative differential mobility was reported in similar cases for passive and active systems.

\section*{Conflicts of interest}
There are no conflicts to declare.

\section*{Acknowledgments}
OC gratefully acknowledges support by the  Lise-Meitner fellowship M 2450-NBL. TF acknowledges funding by  
the Austrian Science Fund (FWF):
P 28687-N27.
The computational results presented have been achieved  using the HPC infrastructure LEO of the University of Innsbruck. 



\balance

\bibliographystyle{rsc}
\bibliography{microswimmers.bib}



\end{document}